\documentclass[a4paper,12pt,english]{article}
\usepackage{amsmath}
\usepackage{graphicx}
\title{\textbf{Theoretical Analysis of Reflection and Refraction of Electromagnetic Waves on an Anisotropic,  Inhomogeneous and Linear Medium}}
\author{\textbf{Andri S. Husein}\footnote{\textit{decepticon1022@gmail.com}},~~~\textbf{Roniyus MS},~~~\textbf{Sri W. Suciyati}\\\\
Department of Physics, University of Lampung\\
Jl. Soemantri Brodjonegoro 1, Bandar Lampung, Indonesia}
\date{}
\begin{document}
\maketitle
\begin{abstract}
Theoretical and computational study of reflection and refraction of electromagnetic waves with $p$-polarized and normal incidence electric wave amplitude vector on anisotropic, inhomogeneous and linear medium has been done.

The medium used in this study is a model of susceptibility in the form of rank-2 tensor which its value depends on the frequency, position and time. Theoretical and computational formulation of transmittance and reflectance are then tested using vacuum and FeF$ _2$ magnetic material.
\end{abstract}

\section{Introduction} 
The interaction between light and matter are interaction between two groups of physical quantities i.e. quantities derived from electromagnetic waves, electric fields, $\vec{E}(\vec{r},t)$, and magnetic fields, $\vec{H}(\vec{r},t)$, with quantities derived from the material i.e. electric susceptibility, $\chi_{e}$, magnetic susceptibility, $\chi_{m}$, and electrical conductivity, $\sigma$. 

In a non-conducting materials, electrical conductivity is zero, so that the material response only expressed by $\chi_{e}$ and $\chi_{m}$. The values $\chi_{e}$ and $\chi_{m}$ in the material are influenced by various factors such as frequency of the incident wave  \cite{azad,kaya,thamizhmani}, temperature \cite{baeraky,krupka}, pressure and porosity \cite{bourlange,donnelly,saar}, inhomogeneous properties  \cite{agron,huttunen,longhi,saville}.

Reflection and refraction of electromagnetic waves in inhomogeneous material are an interesting phenomenon and have several applications, such as: developing radar signal absorbing material \cite{saville}, developing photonic crystals \cite{longhi}, investigations of biological materials  \cite{huttunen}, and investigations of stratified media \cite{agron}. 

Inhomogeneous properties which are used in \cite{agron,huttunen,longhi,saville} derived from the values of materials susceptibility are only changing in spatial coordinates. Research on inhomogeneous materials which are involving changes in susceptibility values by time has not found widely. By using the concept of inhomogeneous medium in space and time, we expect to find more complete picture of reflection and refraction of electromagnetic wave at the surface of anisotropic, inhomogeneous and linear medium.

\section{Medium Properties}
In this study we use medium which is anisotropic, linear, without charge density and inhomogeneous. All types of materials generally have an anisotropic and linear properties except in certain materials such as FeF$_{2}$ \cite{roniyus1,roniyus2,roniyus3,roniyus4,roniyus5} dan GeSe$_{4}$ \cite{patrick}. Both of these anisotropic materials, beside having linear properties, also have nonlinear properties. Material without free charge density is an insulating material. Example of an insulating materials which are commonly used in optical experiments are glass based materials: fluoride glass \cite{lucas}, silica \cite{tan}, soda glass and copper glass \cite{Smith}, selenium glass \cite{patrick} and non-oxide glass \cite{adam}. Some of ceramic based materials are Nd YAG \cite{qu} and NAT \cite{wang}. Some examples of polymer-based materials are polycarbonate (PC) and polyethylene (PET) \cite{yong}.

Mathematically, anisotropic medium is a medium with the values of linear electric susceptibility, $\chi_{e}$, or linear magnetic susceptibility, $\chi_{m}$, has the form of rank-2 tensor (matrix of ordo 3$\times$3). Inhomogeneous medium in this study is a model of electric and magnetic susceptibility tensor that all components are functions of frequency, position and time respectively, have the form $\overleftrightarrow{\chi}_{e}(\omega,r,t)$ and $\overleftrightarrow{\chi}_{m}(\omega,r,t)$ with $ij$ subscript in form tensor components to $ij$. Theoretical and computational studies in this research is done by involving all components (18 entries) susceptibility tensor of the medium to obtain optimal use.

All components of the tensor model above are included in the calculation and will be used for testing a real medium. Considering FeF$_{2}$ magnet materials have been widely used in optics research \cite{roniyus1,roniyus2,roniyus3,roniyus4,roniyus5} and has been qualified as a medium with properties: linear, ansitropic and no charge density, so that the calculation for a real medium will use this material. The tensor form of FeF$_{2}$ are as follow:

\begin{equation}\label{1}
\overleftrightarrow{\chi}_{e}(\omega,r,t)=
\begin{bmatrix}
4.5 & 0 & 0 \\
0 & 4.5 & 0 \\
0 & 0 & 4.5 \\
\end{bmatrix}
\end{equation}
\begin{equation}\label{2}
\overleftrightarrow{\chi}_{m}(\omega,r,t)=
\begin{bmatrix}
\chi_{11}(\omega,r,t) & i\chi_{12}(\omega,r,t) & 0 \\
-i\chi_{21}(\omega,r,t) & \chi_{22}(\omega,r,t) & 0 \\
0 & 0 & \chi_{33}(\omega,r,t) \\
\end{bmatrix}.
\end{equation}

\section{Basic Formulation}
Electromagnetic wave propagation in the medium is shown by Maxwell equations. Maxwell's equations in international units (SI) for no free charge volume density have form as follows \cite{khan}:

\begin{equation}\label{3}
\nabla.\vec{D}(\vec{r},t)=0
\end{equation}

\begin{equation}\label{4}
\nabla.\vec{B}(\vec{r},t)=0
\end{equation}

\begin{equation}\label{5}
\nabla\times\vec{E}(\vec{r},t)+\frac{\partial\vec{B}(\vec{r},t)}{\partial t}=0
\end{equation}

\begin{equation}\label{6}
\nabla\times\vec{H}(\vec{r},t)-\frac{\partial\vec{D}(\vec{r},t)}{\partial t}=0
\end{equation}

where

\begin{equation}\label{7}
\vec{D}(\vec{r},t)=\epsilon \vec{E}(\vec{r},t)
\end{equation}

\begin{equation}\label{8}
\vec{B}(\vec{r},t)=\mu \vec{H}(\vec{r},t)
\end{equation}
with $\epsilon$ and $\mu$ respectively permittivity and permeability of the medium, i.e. medium response when subjected to electric and magnetic fields. Vectors $\vec{E}(\vec{r},t)$ and $\vec{H}(\vec{r},t)$ are respectively the vector amplitude of electric and magnetic waves. 

When the medium of propagation of electromagnetic waves is vacuum, then the value of medium permittivity, $\epsilon$, equal to the vacuum permittivity, $\epsilon_{0}$, and medium permeability, $\mu$, equal to vacuum permeability, $\mu_{0}$. The value of vacuum permittivity, $\epsilon_{0}$, is {8.85$\times$10}$^{-12}$ $C^{2}$/$N.m^{2}$ and vacuum permeability, $\mu_{0}$,  is {4$\pi$$\times$10}$^{-7}$ $T.m/A$. 

When the electromagnetic wave propagates in non vacuum medium, then the value of $\epsilon$ is equal to $\epsilon$ = $\epsilon_{0}$$\epsilon_{r}$, where $\epsilon_{r}$ is the relative permittivity of the medium and medium permeability value, $\mu$, equal to $\mu_{0}$$\mu_{r}$, where $\mu_{r}$ is relative permeability of the medium. This study has used an electric susceptibility, $\chi_{e}$, and magnetic susceptibility,  $\chi_{m}$, of material as response to the presence of electric and magnetic fields.

Linear anisotropic materials have a relative permittivity value of the medium as follows,
\begin{equation}\label{9}
\epsilon^{(r)}_{ii}=1+{\chi}^{(e)}_{ii},~~\epsilon^{(r)}_{ij} = \chi^{(e)}_{ij} 
\end{equation}
and relative permeability value of the medium as follows,

\begin{equation}\label{10}
\mu^{(r)}_{ii}=1+{\chi}^{(m)}_{ii},~\mu^{(r)}_{ij}={\chi}^{(m)}_{ij}
\end{equation}
with superscript $r$, $e$ and $m$ respectively have the meanings $relative$, $electric$ and $magnetic$,  while the subscript $i$ and $j$ in form of matrix components at $ii$  and $ij$. 

Propagation of electromagnetic waves in the medium are expressed by the governing equation of electromagnetic waves in the medium. The formula can be found by put on $curl$ to eq.(\ref{4}) then use eq.(\ref{6}), eq.(\ref{7}) and eq.(\ref{8}) until it is found governing equation of electromagnetic waves in the medium, as follows : 
\begin{equation}\label{11}
\nabla^2\vec{E}(\vec{r},t)-\mu_{0}\frac{\partial^2\vec{D}(\vec{r},t)}{\partial t^2}+\frac{1}{\epsilon_{0}}\nabla[\nabla.\vec{P}(\vec{r},t)]-\mu_{0}\frac{\partial}{\partial t}\left[\nabla\times\vec{M}(\vec{r},t)\right]=0.
\end{equation}
Eq.(\ref{11}) will be required to determine value of the vector wave refraction, $k^{(t)}$, in the medium.

\section{Theoretical and Computational Study of Reflection and Refraction of Electromagnetic Waves} 
Illustration of propagation of electromagnetic waves can be seen in Fig.1 below.
\begin{center}
\includegraphics[scale=0.7]{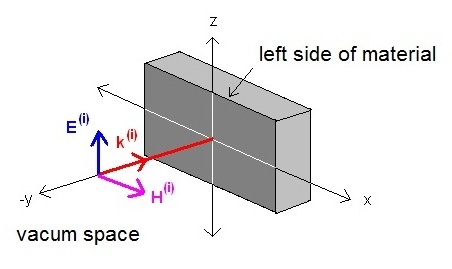}\\
Figure 1. The events of reflection and refraction of electromagnetic waves with $p$-polarized of electric waves incident (located in the y-z plane) at the left boundary surface of the material.
\end{center}
\subsection{The intensity of the incident wave}
Based on Fig.1, it informs that the incident wave vector in vacuum space is
\begin{equation}\label{12}
\vec{k}^{(i)}= k^{(i)}\hat{y}.
\end{equation}
Consider the incident wave comes from vacuum space, so that the incident wave vector is \cite{wangsness}
\begin{equation}\label{13}
k^{(i)}=\omega\sqrt{\epsilon_{0}\mu_{0}}.
\end{equation}
Vector amplitude of the $p$-polarized incident electric wave is defined as follows 
\begin{equation}\label{14}
\vec{E}^{(i)}={E}_{0}\hat{z}.
\end{equation}
Based on eq.(\ref{5}), the vector amplitude of incident magnetic wave is
\begin{equation}\label{15}
\vec{H}^{(i)}=\frac{k^{(i)}}{\mu_{0}\omega}E_{0}\hat{x}.
\end{equation}
The intensity of the incident wave can be found by using the Poynting vector \cite{wangsness}, i.e.
\begin{equation}\label{16}
\langle\vec{S}\rangle=\frac{1}{2}\Re \{\vec{E}(\vec{r},t)\times{\vec{H}}^{*}(\vec{r},t)\},
\end{equation}
then, eq.(\ref{14}) and eq.(\ref{15}) are substituted into eq.(\ref{16}) to obtain the intensity of the incident wave equation
\begin{equation}\label{17}
\langle\vec{S}^{(i)}\rangle=\frac{k^{(i)}{E_{0}}^{2}}{2\mu_{0}\omega}\hat{y}.
\end{equation}

\subsection{The intensity of the reflected wave}
There are two types of reflections that can occur when the electromagnetic waves penetrate the material surface i.e., $ps$ and $pp$-reflection \cite{roniyus3,roniyus4,roniyus5}. Reflection of $ps$, is a reflection of $p$-polarized incident electric waves which generate $s$-polarized reflected electric waves (perpendicular to the plane of incident, i.e. plane of $\textit{x-y}$) with the vector amplitude of the reflected electric wave on $\textit{x}$-axis. Reflection of $pp$, is a reflection of $p$-polarized incident electric waves that generate $p$-polarized of reflected electric waves (parallel to the incident plane, i.e. $\textit{y-z}$ plane) with the vector amplitude of the reflected electric wave on $\textit{z}$-axis. 

The results of theoretical studies using susceptibility tensor of vacuum and FeF$_{2}$ show that, in propagation of light parallel to the $\textit{y}$-axis with $p$-polarized of incident electric waves, the reflectance coefficient of $ps$ has been found equal to zero. Percentage of reflectance totally in the $pp$ reflectance, therefore the $ps$ reflection will not be discussed in further. Reflection of $pp$ produces amplitude vector of reflected electric waves on the $\textit{z}$-axis, so that the vector amplitude of the electric waves is
\begin{equation}\label{18}
\vec{E}^{(r_{pp})}=-r_{pp}E_{0}(\hat{z}).
\end{equation}
where  $r_{pp}$ is a reflection coefficient of $p$-polarized incident electric wave amplitude. 

Based on Snell's Law, the boundary conditions for reflected wave between two different media expressed for the reflected wave vector is
\begin{equation}\label{19}
\vec{k}^{(r_{pp})}=-\vec{k}^{(i)}.
\end{equation}
Furthermore, by using eq.(\ref{5}), eq.(\ref{18}) and eq.(\ref{19}), it can be found amplitude vector for magnetic waves, i.e.
\begin{equation}\label{20}
\vec{H}^{(r_{pp})}=\frac{k^{(i)}}{\mu_{0}\omega}r_{pp}E_{0}(\hat{x}).
\end{equation}
Eq.(\ref{18}) and eq.(\ref{20}) are used to calculate the intensity of reflected waves. The equation of $pp$ reflection intensity for reflected wave is obtained as
\begin{equation}\label{21}
\langle\vec{S}^{(r_{pp})}\rangle=\frac{k^{(i)}{E_{0}}^{2}}{2\mu_{0}\omega}{\mid{r_{pp}}\mid}^{2}\hat{y}.
\end{equation}

\subsection{The intensity of the refraction wave}
The presence of electromagnetic waves on the material surface can raises refraction events \cite{wangsness}. Discussion of wave refraction is started from the relationship between electric field and magnetic field in the material. Eq.(\ref{5}) and eq.(\ref{8}) in forms of the value of electric field related to the magnetic field are
\begin{equation}\label{22} E^{(t)}_{x}=-\frac{\mu_{0}}{k^{(t)}}\bigl(\omega(1+\chi_{33})+i\frac{\partial\chi_{33}}{\partial t}\bigr)H^{(t)}_{z},
\end{equation}

\begin{equation}\label{23} E^{(t)}_{z}=\frac{\mu_{0}}{k^{(t)}}\bigl(\omega(1+\chi_{11})+i\frac{\partial\chi_{11}}{\partial t}\bigr)H^{(t)}_{x}.
\end{equation}
Eq.(\ref{6}) and eq.(\ref{7}) in forms of the wave propagation in the $\textit{y}$-axis direction, so that it can be found the relationship between magnetic field and electric field of the form
\begin{equation}\label{24}
H^{(t)}_{x}=\frac{\epsilon_{0}\omega}{k^{(t)}}\bigl(1+\chi_{e}\bigr)E^{(t)}_{z},
\end{equation}

\begin{equation}\label{25}
H^{(t)}_{z}=-\frac{\epsilon_{0}\omega}{k^{(t)}}\bigl(1+\chi_{e}\bigr)E^{(t)}_{x}.
\end{equation}
If it is assumed that the solution of eq.(\ref{11}) is a harmonic wave function, then by substituting eq.(\ref{22}), eq.(\ref{23})), eq.(\ref{24}) and eq.(\ref{25}) into eq.(\ref{11}), it can be found characteristics matrix  of wave refraction, in the form 
\begin{equation}\label{26}
\begin{bmatrix}
\frac{\mu_{0}\epsilon_{0}}{k^{(t)}}\omega(1+\chi_{e})P+\mu_{0}\epsilon_{0}\omega(1+\chi_{e})R-{k^{(t)}}^{2} & 0 \\
0 & \frac{\mu_{0}\epsilon_{0}}{k^{(t)}}\omega(1+\chi_{e}) M+\mu_{0}\epsilon_{0}\omega(1+\chi_{e})O-{k^{(t)}}^{2}\\
\end{bmatrix}=0,
\end{equation}
where
\begin{equation}\label{27}
P=\frac{\partial}{\partial t}\frac{\partial\chi_{33}}{\partial y}-i\omega\frac{\partial\chi_{33}}{\partial y},
\end{equation}
\begin{equation}\label{28}
R=\omega(1+\chi_{33})+i\frac{\partial\chi_{33}}{\partial t},
\end{equation}
\begin{equation}\label{29}
M=\frac{\partial}{\partial t}\frac{\chi_{11}}{\partial y}-i\omega\frac{\partial\chi_{11}}{\partial y},
\end{equation}
\begin{equation}\label{30}
O=\omega(1+\chi_{11})+i\frac{\partial\chi_{11}}{\partial t}.
\end{equation}
The values of refraction wave vector can be obtained by solving the determinant of characteristics matrix in eq.(\ref{26}) using MATLAB software. The intensity of the electromagnetic wave refraction has form
\begin{equation}\label{31} \langle\vec{S}^{(t)}\rangle=\mu_{0}\epsilon_{e}{\omega}^2(1+\chi_{e})\left|\Re\Bigl\{\frac{t.t^{*}}{k^{(i)}{k^{(t)}}^{*}}\Bigr\}\right|\hat{y},
\end{equation}
\begin{equation}\label{32} \langle\vec{S}^{(r_{pp})}\rangle=\left|\Re\bigl\{-r_{pp}.r_{pp}^{*}\bigl\}\right|\hat{y}.
\end{equation}
Mark of $(^{*})$  in eq.(\ref{31}) and eq.(\ref{32}) means conjugate vector.  Refraction coefficient, $t$, and reflection coefficients, $r_{pp}$, are obtained by equating the tangential components of the amplitude vector at the boundary between two media. The value of $t$ and $r_{pp}$ are as follows:

\begin{equation}\label{33}
t=\frac{2k^{(i)}k^{(t)}}{\mu_{0}\epsilon_{0}{\omega}^2(1+\chi_{e})+k^{(i)}k^{(t)}},
\end{equation}

\begin{equation}\label{34}
r_{pp}=1-t.
\end{equation}
Reflectance and transmittance are calculated as follows :

\begin{equation}\label{35} T=\left|\frac{\langle\vec{S}^{(t)}\rangle.\hat{y}}{\langle\vec{S}^{(i)}\rangle.\hat{y}}\right|,
\end{equation}

\begin{equation}\label{36} R_{pp}=\left|\frac{\langle\vec{S}^{(r_{pp})}\rangle.\hat{y}}{\langle\vec{S}^{(i)}\rangle.\hat{y}}\right|.
\end{equation}
The values of $T$ and $R_ {pp}$ will be visualized with MATLAB software.

\subsection{Transmittance and reflectance in the vacuum and FeF$_{2}$}
To find out the truth of $T$ and $R_ {pp}$ formulation above, it will be tested using vacuum as a first test and then FeF$_2$ magnetic material as a second test material. Based on the law of conservation of energy, in a vacuum, all the energy which arrive should in refraction, while in the non-conducting material, energy which arrive can be reflected and refractived but there is no absorption of energy by the material. 

Given the percentage of the total energy which arrive is 1 (100$\%$), so that the total of false intensity is defined not equal to 1. The results of calculations using MATLAB software for transmittance, $T$, and reflectance, $R_{pp}$, values from vacuum and FeF$_2$ in the influence of external magnetic field 3 Tesla can be seen in Fig.2 below
\begin{center}
\includegraphics[scale=0.27]{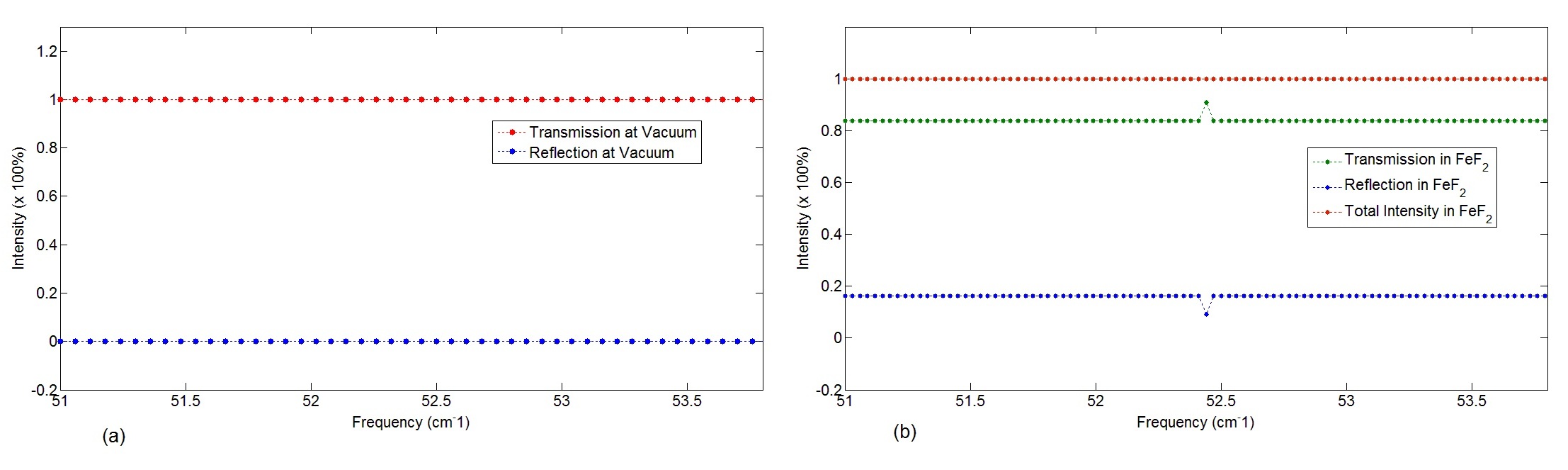}\\
Figure 2. The graph of $T$ and $R_{pp}$ calculation result for (a) vacuum and (b) FeF$_2$ magnetic material with $Total$ as additional graph.
\end{center}
Fig.2 (a) and (b) respectively are result graphs of calculations $T$ and $R_{pp}$ from vacuum and $T$, $R_{pp}$ and $Total$ from FeF$_2$ magnetic materials. 

It is shown in Fig.2 (a) that the value of transmittance curve, $T$, is 1 and $R$ is 0. These results are consistent with the theory of refraction in vacuum. The frequency written in $cm^{-1}$ unit is standard which is used by researchers in the field of optical physics \cite{ritchmyer}.

Fig.2 (b) shows the graph of $T$, $R_{pp}$ and $Total$ calculated from FeF$_2$ magnetic materials. Theoretical study with FeF$_2$ materials generally uses far infrared electromagnetic waves with high intensity in the frequency range 48 cm$^{-1}$ to 58 cm$^{-1}$. In this study, we use the frequency range from 51 cm$^{-1}$ to 53.8 cm$^{-1}$ to shorten the running time because of limitations of computer's processor.

Selection of frequency range from 51 cm$^{-1}$ to 53.8 cm$^{-1}$ is related with material resonance frequency i.e. at 52.45 cm$^{-1}$ \cite{roniyus3}. In Fig.2 (b), it can be seen the value of $T$ and $R_{pp}$ fluctuated around the resonance frequency. Furthermore, it appears the value of $Total$ coincides with value of 1. The $Total$ values which coincide with value of 1 can be used as an indicator that the calculations which are performed using the material FeF$_2$ are correct \cite{roniyus3}.

\subsection{Transmittance and reflectance in an anisotropic, inhomogeneous and linear medium}
Medium with properties of anisotropic, inhomogeneous and linear used in this study is a model of susceptibility tensor as shown in eq.(\ref{1}) and eq.(\ref{2}) with shape of susceptibility curve similar to the curve of FeF$_2$ magnetic susceptibility at far infrared frequency. 

Consider that the FeF$_2$  susceptibility curve has similar to the fluctuations tangent function. After the modification, it can be obtained tangent function approaches the curve of FeF$_2$ magnetic materials susceptibility under the influence of external magnetic field of 1 Tesla, which is
\begin{equation}\label{37}
\chi_{ij}(\omega,r,t)=\rho(r,t)\beta_{ij}\tan\left(\frac{\zeta(r,t)-82.26}{\omega}\right),
\end{equation}
with $\beta_{11}=0.7/8.55\times10^{\pi+3}$, $\beta_{12}=\beta_{21}=1/8.55\times10^{\pi+1.46}$, $\beta_{22}=\beta_{33}=1/8.55\times10^{\pi}$.
Factor of $\rho(r,t)$ and $\zeta(r,t)$ in eq.(\ref{37}) is the homogeneity parameter of materials. Homogeneity parameter of materials is defined with:\\
$$\vbox{\offinterlineskip
\halign{\strut \quad $#$\quad & \hfil \quad #\quad \hfil & \quad $#$\quad & \hfil \quad #\quad \hfil  \cr
\noalign{}
$Homogeneous$ & when\;$\rho(r,t)=c$, $c\in\Re$ & $when$\;\zeta(r,t)=c, \;c\in\Re,\cr
\noalign{}
$Inhomogeneous$ & when\;$\rho(r,t)=\rho^{L}(r)\pm\rho^{L}(t)$ &$when$\;\zeta(r,t)=\zeta^{L}(r)\pm\zeta^{L}(t).\cr
\noalign{}
}}$$
The sign of ($..^{L}$) means linear. The figures below are visualization of eq.(\ref{37}) when the medium is homogeneous $\rho(r,t)=1$, $\zeta(r,t)=0$ and magnetic susceptibility FeF$ _2$ in the influence of external magnetic field of 1 Tesla \cite{roniyus3} :

\begin{center}
\includegraphics[scale=0.27]{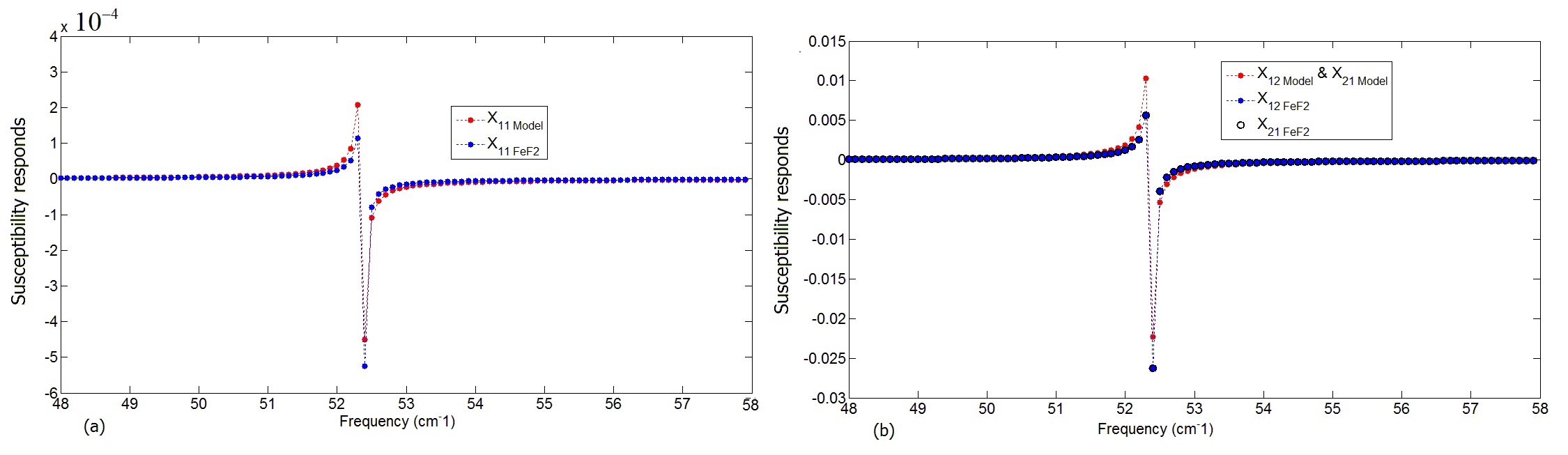}\\
\includegraphics[scale=0.27]{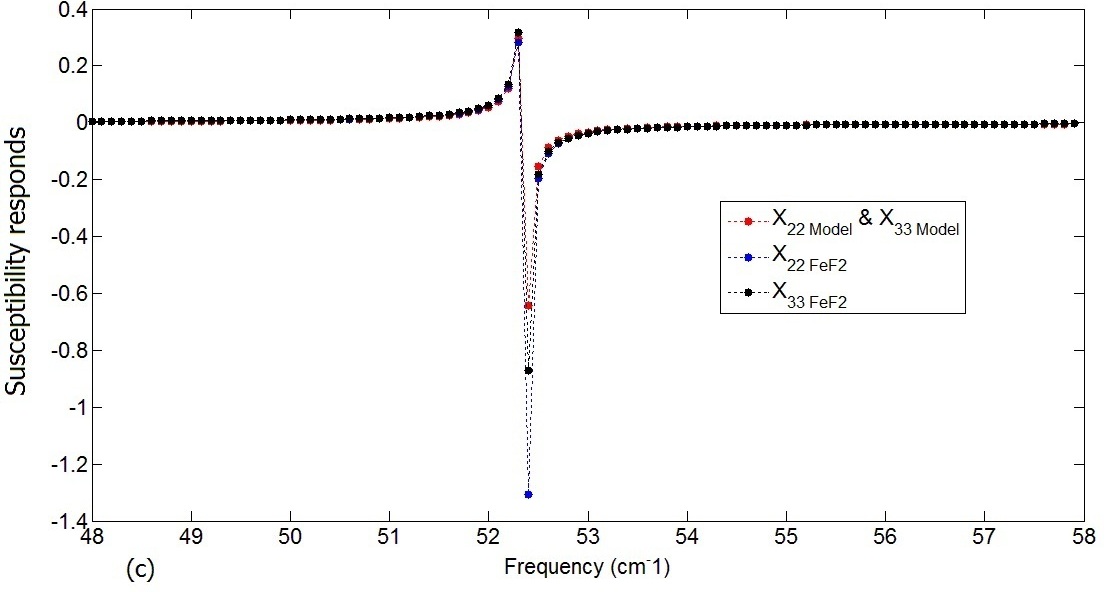}\\
Figure 3. Magnetic susceptibility curves of models and FeF$_2$ appears to fluctuate around the frequency 52.45 cm$^{-1}$, i.e. the resonance frequency of FeF$_2$. (a) $\chi_{11}$, model (red) and $\chi_{11}$, FeF$_2$ (blue), (b) $\chi_{12}$, $\chi_{21}$, models (red) and $\chi_{12}$, $\chi_{21}$, FeF$_2$ (blue and black), (c) $\chi_{22}$, $\chi_{33}$, model (red) and FeF$_2$ (blue and black).
\end{center}

Based on eq.(\ref{37}) note that the value of the magnetic susceptibility is influenced by two homogeneity parameters of the medium i.e., $\rho(r,t)$ (rho) and $\zeta(r,t)$ (zeta). Formulation the value of $\rho(r,t)$ and $\zeta(r,t)$, to simplify the problem, can use Agron and Gogineni approach on stratified materials \cite {agron}. 

In stratified materials, review is performed in the direction of light propagation. Given the direction of light propagation in this study is $y$-axis, then it can be selected $\rho(r,t)=1-y$, $\zeta(r,t)=0$ and $\rho(r,t)=1$, $\zeta(r,t)=y$ for an example of calculation. The calculation results of $T$ and $R_{pp}$ values using these parameters at the point (0, 0, 0), are shown in Fig.4 as follows 
\begin{center}
\includegraphics[scale=0.27]{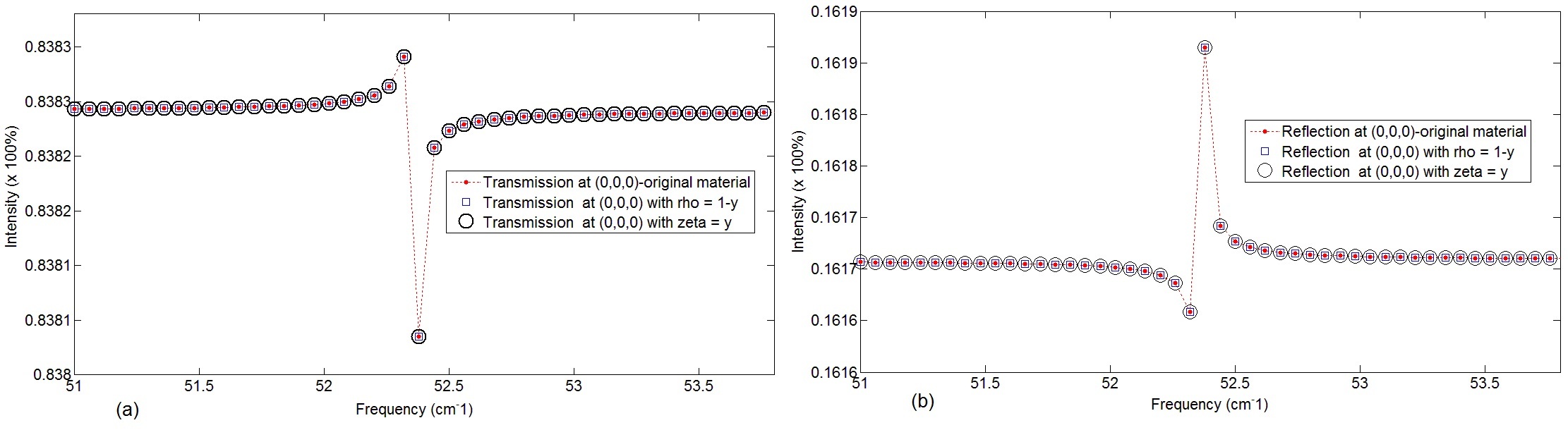}\\
Figure 4. (a) The comparison values of $T$ and $R_{pp}$ of the model depend on the position and (b) depend on the position of the magnetic susceptibility at the origin (0,0,0). 
\end{center}

Fig.4 shows the effect of inhomogeneous magnetic susceptibility values in the direction of light propagation. The values of transmittance is shown in Fig.4 (a), whereas the reflectance values is shown in Fig.4 (b). Both figures show that there are no change in the pattern of transmittance and reflectance from initially state. 

The parameters of inhomogeneous by time can be expressed by functions $\rho(r,t)=1-t$, $\zeta(r,t)=0$ and $\rho(r,t)=1$, $\zeta(r,t)=t$. The results of calculation of $T$ and $R_{pp}$ using linear functions at the point (0, 0, 0) and $t$ = 0, 0.5 $s$ are shown in Fig.5 as follow
\begin{center}
\includegraphics[scale=0.27]{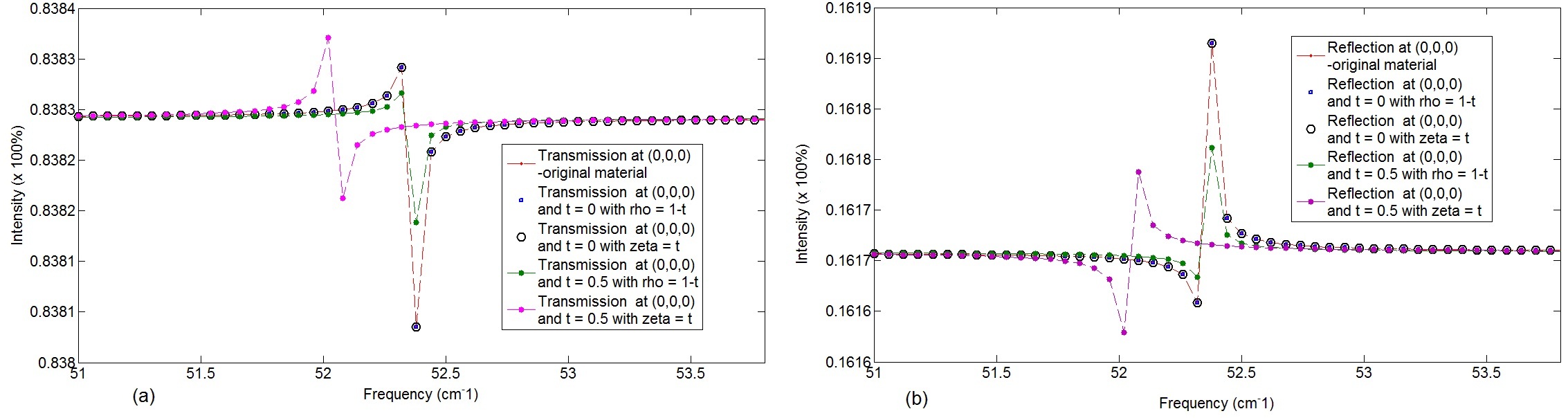}\\
Figure 5. The comparison values of $T$ and $R_{pp}$ from the magnetic susceptibility model of time dependent and time independent at the origin (0, 0, 0) when $t$ = 0, 0.5 $s$. (a) The values of $T$ on $t$ = 0 $s$ (blue rectangle and black circle), $t$ = 0.5 $s$ (green dot and pink dot), (b) The value of $R_{pp}$ on $t$ = 0 $s$ (blue rectangle and black circle), $t$ = 0.5 $s$ (green dot and pink dot). 
\end{center}

Fig.5 shows the effect of inhomogeneous magnetic susceptibility values by time. Transmittance value, $T$, is shown in Fig.5 (a), whereas the reflectance value $R_{pp}$ is shown in Fig.5 (b). Both curves, $T$ and $R_{pp}$, (blue rectangle and black circle) in the beginning ($t$ = 0 $s$), coincide with transmittance and reflectance curves of the medium which are independent of time. But in further time ($t$ = 0.5 $s$), the fluctuations of $T$ and $R_{pp}$ (green dot and pink dot) are decreased and shifted relative to its original state. Changes in $T$ and $R_{pp}$ at inhomogeneous medium thus can be generated by the parameters of time. 

\section{Conclusions} 
Computational results using vacuum susceptibility shows that the value of transmittance, $T$, is 1 ($ 100 \% $), the reflectance, $R_ {pp}$, is 0 ($ 0 \% $) and the total intensity is 1 (100 $ \% $), according to the theory of reflection and refraction of electromagnetic waves that propagate in vacuum. Computational results of FeF$ _2$ magnetic material shows that the total value of intensity is 1 (100 $ \% $), according to the results of previous studies. 

The results of theoretical and computational studies on the surface of an anisotropic, inhomogeneous and linear medium shows that the values of transmittance, $T$, and reflectance, $R_ {pp}$, are not affected by variations of susceptibility values in the direction of light propagation. The changes in values of transmittance, $T$, and reflectance, $R_ {pp}$, on the medium surface can be caused by time.

In other words, in the direction of light propagation ($y$-axis), the values of transmittance, $T$, and reflectance,  $R_{pp}$, in the surface of medium are not affected by the changes of magnetic susceptibility values �as shown in Fig.4 (a) and (b). Variations in medium susceptibility values by time can change the fluctuation of transmittance, $T$, and reflectance, $R_{pp}$, at the surface of medium by two mechanisms as shown in Fig.5 (a) and (b).

\section{Acknowledgments}
ASH thank to Computational Laboratory, University of Lampung for supporting software and hardware. Thank to all of Physics Department Staffs, colleagues, friends whom I am not able to mention one by one. Thank for everything. In depth, ASH thank to beloved Mother for sincere and limitless love. Aliya Syauqina Hadi for her cuteness and cheerfulness.

\end{document}